\documentclass[nohyper,11pt,letterpaper]{JHEP3}
\usepackage{graphicx}

\DeclareSymbolFont{AMSa}{U}{msa}{m}{n}
\DeclareSymbolFont{AMSb}{U}{msb}{m}{n}
\let\Box\relax
\DeclareMathSymbol{\Box}{\mathord}{AMSa}{"03}

\def \eqn#1#2{\begin{equation}#2\label{#1}\end{equation}}

\title{Evolution of Gravitationally Unstable de Sitter Compactifications}

\author{C. Krishnan, S.Paban, M. \v{Z}ani\'{c}\\
Department of Physics \\ University of Texas, Austin,
TX 78712}

\abstract{We study the time evolution of unstable $dS_p \times S^q$ configurations with flux in  theories of gravity with a cosmological constant. For certain values of the flux, we identify a stable configuration to which these unstable solutions flow. For other values of the flux the sphere wants to decompactify,  regardless of the sign of the initial perturbation. }

\received{???????? ?st, 2000} \accepted{???????? ?th, 1998}
\preprint{\hepth{0503025 }\\UTTG-03-05}

\begin{document}



\section{\bf Introduction}
It has been known since the work of Freund and Rubin \cite{Freund:1980xh} that when an antisymmetric tensor field of dimension  $q-1$ is added to gravity (Einstein-Hilbert action), space-times of dimension $p+q$  can naturally compactify to product spaces of the form $dS_p \times S^q$.

Bousso, DeWolfe and Myers \cite{Bousso:2002fi}
derived analogous  results when a  positive cosmological constant is added to the gravitational action, and studied the stability of these compactifications under
gravitational perturbations. Their  analysis  showed the stability
to depend on the relative value of the flux compared to the cosmological constant as well
as on the dimension of the internal space.

The unstable modes, at these compactifications, were identified in \cite{Bousso:2002fi},  generalizing the work of \cite{DeWolfe:2001nz}. In this paper,
we  derive
their equations of motion and study their evolution. In doing so we will make the assumption that these
remain the only unstable modes throughout the evolution. 
A similar analysis, in the absence of flux,  was recently done by  Contaldi, Kofman and Peloso \cite{Contaldi:2004hr}.

This paper answers the questions about the  fate of these configurations that was posed in the original paper of Bousso, DeWolfe and Myers.  It is unclear to what extent the lessons learned here translate into more realistic models, e.g. \cite{Kachru:2003aw,Saltman:2004jh}.  In fact,  Giddings and Myers  \cite{Giddings:2004vr}  have already studied these types of models and argued that positive vacuum energy together with extra dimensions render the four-dimensional space unstable toward decompatification of the extra dimensions. Our study is purely classical and does not incorporate the effect of thermal fluctuations or tunneling considered in \cite{Giddings:2004vr}.


After this work was completed, we learned that  the time evolution of these configurations had already  been studied in the past \cite{Okada:1984cv,Kolb:1986nj}. Our results present a more complete analysis of all the initial conditions and agree
with them when these conditions overlap. The goal of these earlier papers was to use the unstable mode as the inflaton mode.
Other references that partially overlap with this work are \cite{Navarro:2004mm},\cite{Lu:2003dn},\cite{Lu:2003iv}.

In section 2 we present a summary of the solution described in \cite{Bousso:2002fi} to establish notation and state the problem. In section 3, we display a more general solution of the equations of motion that reduces to the solutions found in \cite{Bousso:2002fi} and hence can interpolate
between an unstable initial configuration and a stable final state. In section 4, we will present the time evolution that results
from the numerical analysis.

\section{\bf Product Spacetimes with Flux}

We will consider solutions of the following action;

\eqn{action}{ S= \frac{1}{2} \int d^{p+q} x \sqrt{-g} ( R - 2 \Lambda - \frac{1}{2 q!} F_{P_1...P_q} F^{P_1...P_q} )}
in units where $M_P$ in $ p+q$ dimensions has been set equal to 1.  The cosmological constant $\Lambda$ will be assumed positive. $F_{P_1...P_q} $ is a totally antisymmetric tensor of rank $q$. Throughout this paper we adopt the notation that upper case latin indices run from 1 to $ p+q$,
greek indices run from $1$ to $p$ and lower case latin indices run from $1$ to  $q$. The equations of motion are

\eqn{EOMg}{ G_{MN} =  \frac{1}{2 (q-1)!} F_{M P_1...P_{q-1}} F_N^{P_1...P_{q-1}} - g_{MN} \,\, (\Lambda + \frac{1}{4 q!} F_{P_1...P_q} F^{P_1...P_q})  }
\eqn{EOMA}{\frac{1}{\sqrt{-g} }\partial_M ( \sqrt{-g} F^{M P_1...P_{q-1}} ) =0 }
These equations admit a solution that is a product of two spaces, a Lorentzian and a Riemannian ($S^q$):

\eqn{metric} { ds^2 = -dt^2 + a(t)^2 d^{p-1}{\bf x} + R_0^2  d\Omega_q}

\eqn{flux}{F_q= c \, \mbox{vol}_{ S^q} \hspace{5ex}  \oint \mbox{vol}_{ S^q} = R_0^q \,\Omega_q}
To be consistent with the notation of \cite{Bousso:2002fi} we will parametrize the solutions in terms of

$${\cal F} = \frac{c^2}{ 4 \Lambda} $$ Though ${\cal F}$ has been often referred to, in the literature, as the flux, it is clear from (\ref{flux}) that the actual flux, with the normalization given above, is also a function of the compactification radius $R_0$. With  a normalization that makes the
flux radius independent, $F_q= \frac{c}{R^{q} (t)}  \, \mbox{vol}_{ S^q}$, there are
actually two static solutions, instead of one,  for some values of $c$. 
The equations of motion for the Lorentzian scale factor, $a(t)$ are:

\begin{eqnarray}
 \frac{(p-1)(p-2)}{2} \frac{\dot{a}^2}{a^2} +  \frac{(p-1)(p-2)}{2} \frac{k}{a^2} &= &\Lambda + \Lambda \, {\cal F}-\frac{q(q-1)}{2}\frac{1}{R_0^2} \nonumber\\
  \label{static} \end{eqnarray}
 $k=1,0,-1$ corresponds to closed, flat or open spaces.
The effective cosmological constant of the $p$ dimensional Lorentzian space, $\Lambda_p$, is given by:
\eqn{Leffec}{ (p-1) \Lambda_p= \frac{2 \Lambda}{p+q-2} \left( 1 - ( q-1) {\cal F} \right) }
Since $\Lambda$ has been chosen positive, the effective cosmological constant  will be positive as long as

\eqn{bflux}{0 \leq {\cal F } < \frac{1}{q-1}}
The radius of $S^q$ is given by:

\eqn{radius} {\frac{q-1} {R_0^2} = \frac{ 2 \Lambda}{p+q-2} \left( 1 + (p-1) {\cal  F }\right)}
The following table summarizes the regions of stability as a function of the free dimensionless parameter ${\cal F}$ and the dimension of the Riemannian space $q$ as found in \cite{Bousso:2002fi}.

\vspace{3ex}
\begin{tabular} {|c|c|c|c|c|} \hline
          &  \multicolumn{2}{c|}  {$dS_p \times S^q$ } & Minkowski $\times S^q$ &  $AdS_p \times S^q$  \\  \hline
 $q$ &   {\em unstable}  &   {\em stable}  &  &    \\  \hline
   2    &    $0  \leq  {\cal F}  < \frac{1}{(p-1)} $ &  $\frac{1}{(p-1)}  \leq {\cal F } < 1 $ &
   $ {\cal F} = 1$ &  $   1   < {\cal F} $ \\  \hline
   3      &    $0  \leq  {\cal F}  < \frac{1}{2(p-1)} $ &  $\frac{1}{2(p-1)}  \leq {\cal F } < \frac{1}{2}$ &
  $ {\cal F} = \frac{1}{2}$ &  $   \frac{1}{2} < {\cal F} $ \\  \hline
   4      &   $0  \leq  {\cal F}  < \frac{1}{3(p-1)} $, $\frac{1}{2 (p-1)} < {\cal F} < \frac{1}{3}$   &  $\frac{1}{3(p-1)}  \leq {\cal F } < \frac{1}{2(p-1)}$  & ${\cal F} = \frac{1}{3}  $ &   $   \frac{1}{3} < {\cal F} $ \\  \hline
$  \geq  5$  &      $ 0 \leq {\cal F}  < \frac{1}{q-1} $ &    --  &   ${\cal F }=\frac{1}{q-1}$ &  $\frac{1}{q-1} < {\cal F}$ \\ \hline

\end{tabular}

\vspace{3ex}

Not reflected in the previous table is the fact that the number of unstable modes increases 
as a function of $q$. For $q=2, \mbox{and} \,\, q=3$ there is only one unstable mode. It corresponds
to the radius  of $S^q$. For $q \geq 4$ other modes became tachyonic as well.

\section{\bf Evolution of Unstable Spacetimes}
There is a broader set of solutions to (\ref{EOMg}),(\ref{EOMA}), that do not correspond to a product spacetime. They are of the form:

\eqn{metric} { ds^2 = -dt^2 + a(t)^2 d^{p-1}{\bf x} + R(t)^2  d\Omega_q}

\eqn{flux0}{F_q= \frac{f}{R^{q} (t)}  \, \mbox{vol}_{ S^q} \hspace{5ex}  \oint \mbox{vol}_{ S^q} = R(t)^q \,\Omega_q} In this case, the normalization of the differential form $F_q$ is fixed by the equations of motion. 
The corresponding equations of motion are:

\begin{eqnarray}
\frac{q(q-1)}{2 R^2}(1+ \dot{R}^2) + \frac{(p-1)(p-2)}{2} \frac{\dot{a}^2}{a^2} + q(p-1) \frac{\dot{R}\,\dot{a}}{R \,a} &= &\Lambda \left(1 + \frac{\beta }{\Lambda ^q R ^{2q}}\right)  \nonumber\\
\frac{q(q-1)}{2 R^2} ( 1 + \dot{R}^2 )+ \frac{(p-2)(p-3)}{2}\frac{\dot{a}^2}{a^2}+q(p-2) \frac{\dot{R} \, \dot{a}}{R \, a} + (p-2) \frac{\ddot{a}}{a} + q \frac{\ddot{R}}{R}  &= &\Lambda\left(1 + \frac{\beta }{\Lambda ^q R ^{2q}}\right) \nonumber \\
 \frac{(q-2)(q-1)}{2 R^2}(1+ \dot{R}^2)+\frac{(p-1)(p-2)}{2} \frac{\dot{a}^2}{a^2} +  (q-1)(p-1) \frac{\dot{R} \, \dot{a}}{R \,a} +  (p-1)\frac{\ddot{a}}{a} +(q-1) \frac{\ddot{R}}{R} & = &\Lambda \left(1 - \frac{\beta }{\Lambda ^q R ^{2q}}\right) \nonumber \\
  \label{EOM} \end{eqnarray}
where $\frac{\beta}{ \Lambda^q} = \frac{f^2}{4 \Lambda}$. To lighten these already complicated equations we have made 
the assumption that $k=0$, the generalization to $k=\pm1$ is straightforward. These equations can be reduced into an equation for the breathing mode of the sphere, often called the radion: $R(t)$

\eqn{Rmotion}{ \ddot{R} = \frac{q+p-2}{p-2}  \frac{ \dot{R}^2}{R}  \mp \frac{p-1}{p-2} \dot{R} \sqrt{ \frac{q(p+q-2)}{p-1} \frac{ \dot{R}^2}{R^2} - 2 \frac{p-2}{p-1} \left[ \frac{q(q-1)}{2 R^2} - \Lambda \left( 1 + \frac{\beta} {\Lambda^q R^{2 q}}\right)\right]} -\frac{dV(R)}{dR}}
where the potential

\eqn{potential}{ V(R) = -\frac{ \Lambda R^2 }{ p+q-2} + \frac{ (p-1)  }{ (q-1)(p+q-2)} \frac{\beta}{ (\Lambda R^2 )^{q -1} }  +(q-1) \ln (R)}
The evolution of $a(t)$ is given by:
\eqn{evolutiona}{\frac{\dot{a}}{a}= -\frac{q}{p-2} \frac{\dot{R}}{R} \pm \frac{1}{p-2} \sqrt{ \frac{q(p+q-2)}{p-1} \frac{ \dot{R}^2}{R^2} - 2 \frac{p-2}{p-1} \left[ \frac{q(q-1)}{2 R^2} - \Lambda ( 1 + \frac{\beta}{ \Lambda^q  R^{2 q}})\right]}} 
The initial conditions, i.e. the sign of $\frac{\dot{a}}{a}$ at $\dot{R}=0$,  and $R$ equal  the radius of the unstable point,
decide which branch of the equations (\ref{Rmotion}), (\ref{evolutiona}) to consider. 

\begin{figure}
\includegraphics{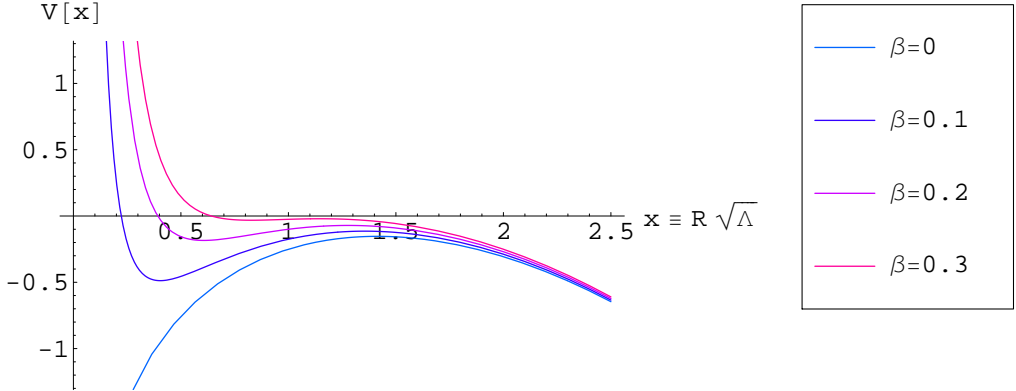}
\caption{ Potential for the field $R(t)$ as a function of the flux, $\beta$, when  $p=4$,  and $q=2$.}
\end{figure}
As long as the flux is nonvanishing and relatively small, $ 0 <  \beta_s $, where
\eqn{beta_s}{\beta_s=\frac{(q-1)^{2q-1}}{p-1}\left(\frac{p+q-2}{2q}\right)^q}
the radion potential  will  always have  two stationary  solutions, one stable and one unstable.   There is a solution to the equations of motion  that corresponds to the field $R(t)$ sitting
on the maximum. In this case,  the geometry factorizes and corresponds to a $dS_p \times S_q$ space.  There are also solutions that correspond to $R(t)$ sitting in the minimum. In these cases, the corresponding
geometry also factorizes into the product of the spaces $L_p \times S_q$. The Lorentzian
signature space will be a (A)dS or flat Minkowski depending on the value of $\beta$. In particular, for $ \beta_m < \beta < \beta_s $, where
\eqn{beta_m}{\beta_m=\frac{(q-1)^{2q-1}}{2^q}}
the Lorentzian signature space will also be a de Sitter space.

We are interested in studying the evolution of the unstable solutions away from the maximum.
The equations (\ref{Rmotion}),(\ref{evolutiona}) will capture such an evolution
provided that $R(t)$ is indeed the only mode excited along the path away from the maximum.
The analysis of \cite{Bousso:2002fi} proves that this is indeed the only excited mode around the
extrema of the potential, provided that $q=2, 3$, for  $q \geq 4$  there
will be other unstable  modes. In this work we will restrict our attention to the first two cases, and leave the analysis of  $q \geq 4$ to future work.

\begin{figure}
\includegraphics{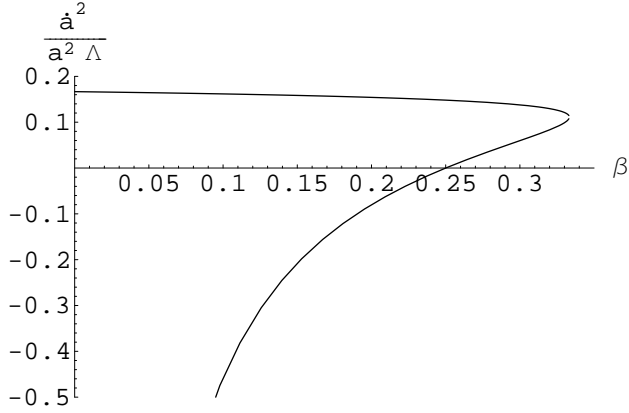}
\caption{Value of  $\frac{\dot{a}^2}{a^2} $ at the stationary points as a function of the initial condition $\beta$. The upper branch corresponds to the unstable point and the lower branch corresponds to the stable point. (Case $p=4$,  $q=2$.)}
\end{figure}

\section{\bf Results}

In this section we report on the numerical solution of  the equations (\ref{Rmotion}), (\ref{evolutiona}). It is useful to distinguish between two cases:

\begin{itemize}
\item When  the minimum of the potential corresponds to another de Sitter solution ($ \beta_m < \beta < \beta_s$)

\item When the minimum of the potential corresponds to an Anti de Sitter solution ($ 0 < \beta < \beta_m$)
\end{itemize}\
\vspace{2ex}

\subsection{$ \bf{\beta_m < \beta < \beta_s}$}
\vspace{1ex}

\begin{figure}
\includegraphics{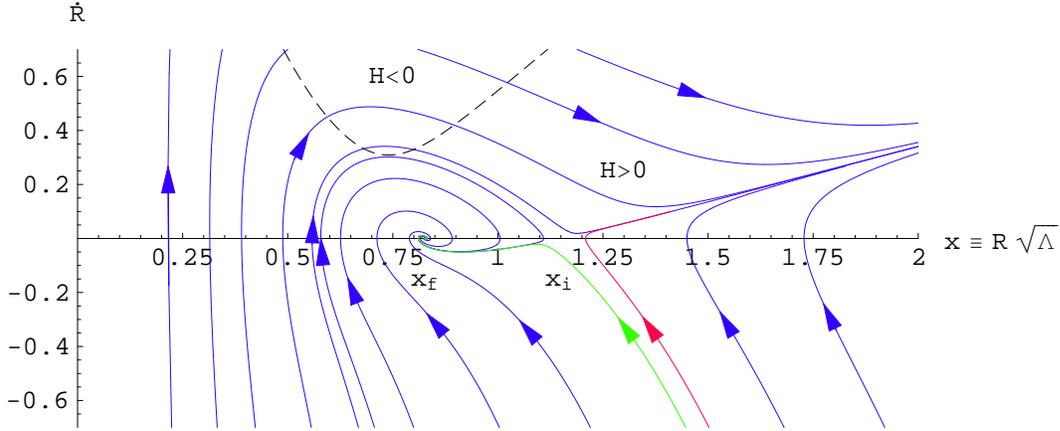}
\caption{ Numerical evolution for $p=4$, $q=2$, $\beta = 0.3$.  The arrows indicate the direction of time evolution.}
\label{fig:dS2dS}
\end{figure}

Let us choose the initial condition such that $\dot{a} /a >0$, namely, we start with an expanding de Sitter space. The numerical analysis shows that the evolution from the unstable  $dS_p \times  S^q$ in this case leads either to a decompactification of the compact dimensions, or to another, stable $dS_p \times S^q$ solution, see Figure \ref{fig:dS2dS}.

In the case of decompactification, the expansion rate of the radius of the sphere $S^q$ asymptotes to:

\eqn{asymptotic_expansion}{\frac{\dot{R}}{R}=\sqrt{\frac{2 \Lambda}{(D-1)(D-2)}}}
where $D=p+q$. It is interesting to note that this expansion rate matches the asymptotic effective Hubble constant for the de Sitter space, i.e. $\dot{R}/R=\dot{a}/a$ as $R,a \rightarrow \infty$.
This agrees with the result found in \cite{Contaldi:2004hr},  because this fast evolution erases all
the information about the initial flux.

The other direction of evolution from the unstable $dS_p \times  S^q$, with these initial conditions, leads to a stable  $dS_p \times  S^q$ solution. The two solutions are characterized by the same value of $\beta$. Namely, for a static solution we have:

\eqn{beta_st}{\beta / \Lambda^q={\cal F} R^{2q}}

\begin{figure}
\includegraphics{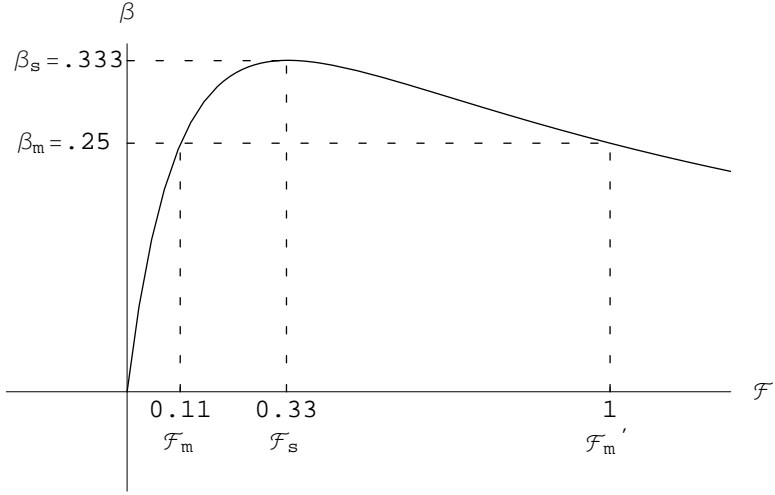}
\caption{Flux $\beta$ as a function of parameter ${\cal F}$, when $p=4$ and $q=2$.}
\end{figure}

\eqn{F_st}{R^2=\frac{(q-1)(D-2)}{2\Lambda [1+(p-1){\cal{F}}]}}
For a particular value of $\beta$ we have two static solutions related by:

\eqn{duality}{\frac {{\cal F}_f}{{\cal F}_i}=\left (\frac{1+(p-1){\cal
F}_f}{{1+(p-1){\cal F}_i}}\right)^q} The stable solution will always have
a smaller radius $R$ than the initial, unstable one. The two are
related by:

\eqn{Rduality}{R_f^2=\frac{1+(p-1){\cal F}_i}{1+(p-1){\cal F}_f}R_i^2}
The effective cosmological constant  for the de Sitter is given by (\ref{Leffec}), and it therefore changes:

\eqn{Hduality}{\left(\frac{\dot{a}}{a}\right)_f^2=\frac{1+(1-q){\cal F}_f}{1+(1-q){\cal F}_i}\left(\frac{\dot{a}}{a}\right)_i^2}

In particular, for $p=4$ and $q=2$ we have:

\eqn{f42}{{\cal F}_f=\frac{1}{9 {\cal F}_i}}

\eqn{R42}{R_f^2=\frac{2}{\Lambda}-R_i^2 = 3 {\cal F}_i R_i^2}

\eqn{H42}{\left(\frac{\dot{a}}{a}\right)_f^2=\frac{(9{\cal F}_i-1)}{9{\cal F}_i (1-{\cal F}_i)}\left(\frac{\dot{a}}{a}\right)_i^2}

\eqn{Lambda4}{\Lambda_{4f}=\frac{\Lambda_{4i}-\frac{4}{27}
\Lambda}{6\Lambda_{4i}-\Lambda} \Lambda}

\begin{figure}
\includegraphics{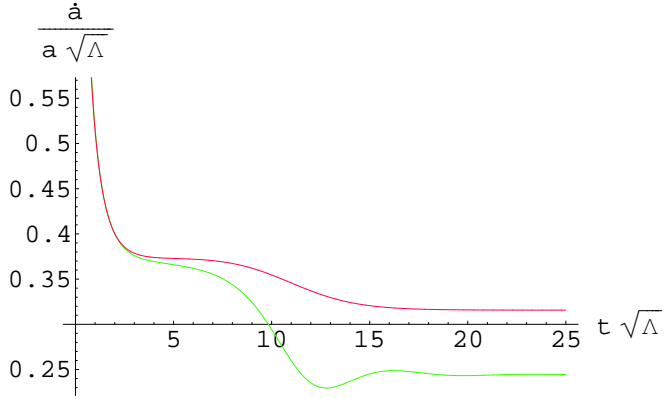}
\caption{The evolution of effective Hubble rate for $p=4$, $q=2$. (The colors match the numerical evolution in Figure \ref{fig:dS2dS})}
\end{figure}

Notice, that the final effective  cosmological constant  can be made very small compared with the initial. An analogous transition has been found in the context of braneworlds \cite{Martin:2003yh}. The radius of the internal sphere will remain, however, within the same order of magnitude of the initial one.  Let us compare the entropies of the initial and the final
solution:

\eqn{entropy}{\frac{S(dS_4 \times S^2)_f}{S(dS_4 \times 
S^2)_i}=\frac{A(dS_4)_f \times V(S^2)_f}{A(dS_4)_i \times 
V(S^2)_i}=\frac{(1-{\cal F}_i)(1+3{\cal F}_i)}{(1-{\cal F}_i)(1+3{\cal F}_i)}=27({\cal F}_i)^2
\frac{1-{\cal F}_i}{9{\cal F}_i -1}>1}
as long as, $\frac{1}{9}< {\cal F}_i < \frac{1}{3}$, which corresponds to
our initial unstable static solution.

\begin{figure}
\includegraphics{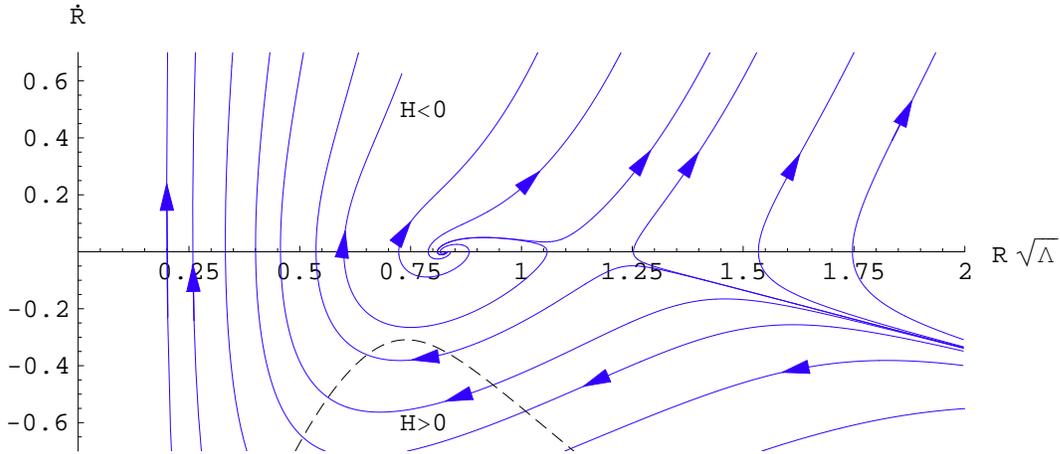}
\caption{ Numerical evolution for $p=4$, $q=2$, $\beta = 0.3$ }
\label{fig:dS2dScontracting}
\end{figure}
\begin{figure}
\includegraphics{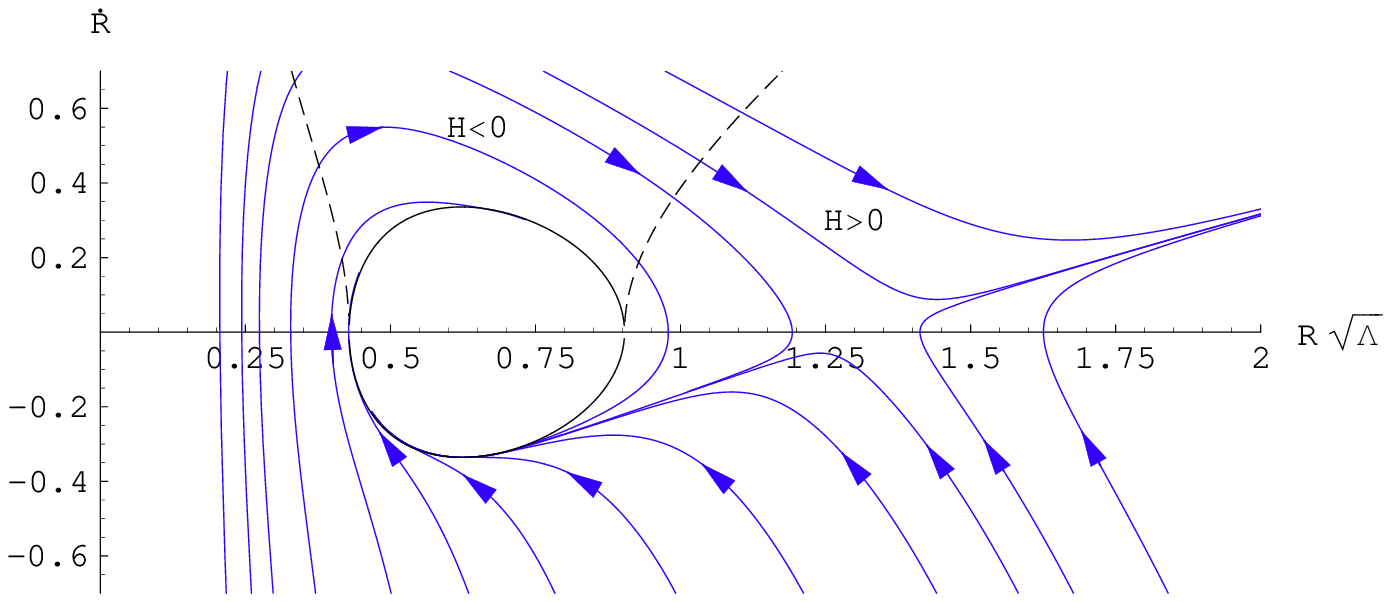}
\caption{ Numerical evolution for $p=4$, $q=2$, $\beta = 0.15$ or equivalently Anti de Sitter minimum.}
\label{fig:AdSminexpanding}
\end{figure}
\begin{figure}
\includegraphics{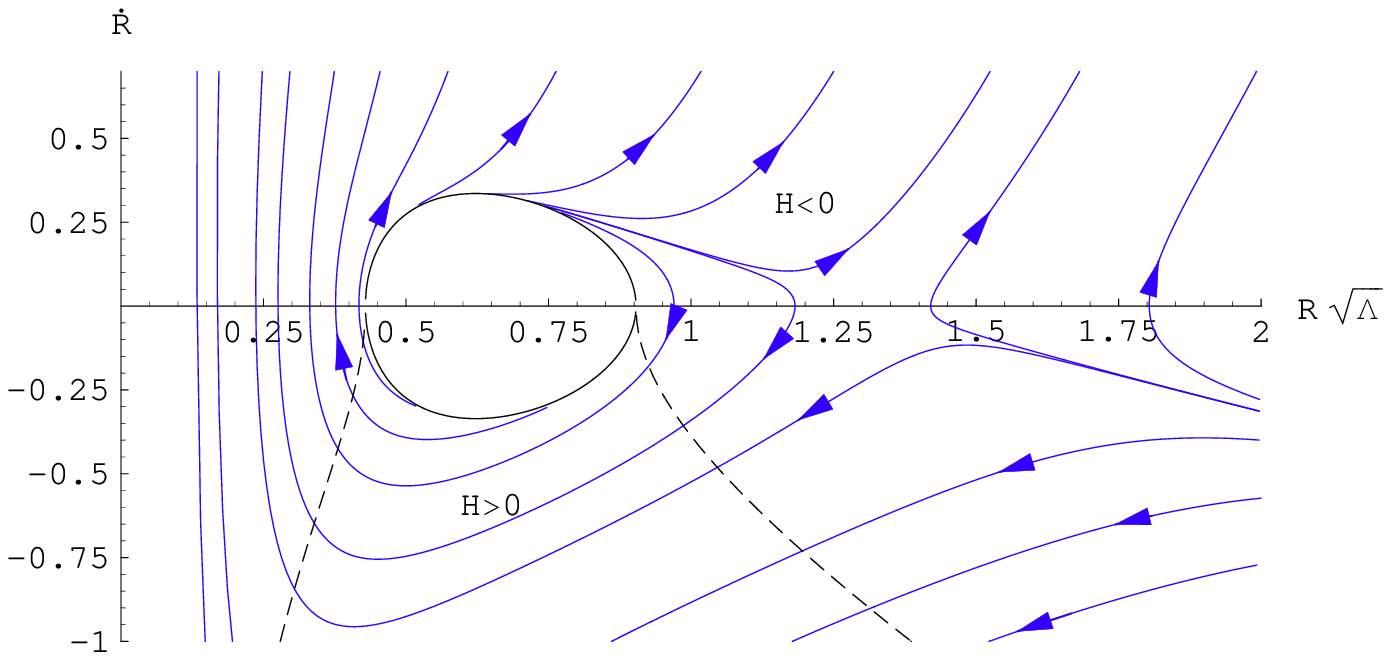}
\caption{ Numerical evolution for $p=4$, $q=2$, $\beta = 0.15$ or equivalently Anti de Sitter minimum. }
\label{fig:AdSmincontracting}
\end{figure}

Now let's look at the situation where the initial conditions are chosen such that the de Sitter component of the space is initially contracting, $\dot{a} /a <0$ (See Figure \ref{fig:dS2dScontracting}). In this case, the numerical evolution shows that the internal dimensions necessarily decompactify, while the initial de Sitter dimensions keep contracting.  This result may appear surprising, at first, when the initial velocity 
at the unstable point is taken to be negative, i.e. $\dot{R}<0$.  These initial conditions choose 
the negative branch in (\ref{evolutiona}) and hence the positive branch in  (\ref{Rmotion}). The friction force in (\ref{Rmotion})  is then big enough to overcome the force generated by the potential. 

The crunching solution should (\ref{metric})  asymptotically become a Kasner type solution.
We did not use this ansatz in the numerical analysis that led to Figure \ref{fig:dS2dScontracting}.

\vspace{2ex}

\subsection{$ \bf{0< \beta < \beta_m}$}
\vspace{1ex}

When the flux $\beta$ is very small, only one of the static solutions corresponds to a de Sitter compactification, the other solution corresponds to an anti de Sitter space. In this case the numerical solution in both branches (either initially contracting (Figure \ref{fig:AdSmincontracting}), or expanding (Figure \ref{fig:AdSminexpanding}) de Sitter phase) leads to a decompactification of the inner space, while the initially de Sitter components either expand or contract, depending on the initial conditions and the initial direction of the numerical evolution. This is a sensible result. The classical trajectories studied here
cannot connect spaces with different spacial curvatures. Given the ansatz for the metric  (\ref{metric}) the initial de Sitter configuration can be 
either flat or closed while the final anti de Sitter configuration has to be open.

\section{\bf Conclusions}


In this paper we have studied the evolution of gravitationally unstable $dS_p \times S^q$  compactifications. We have identified an additional static stable solution that only exists for a certain interval of fluxes.  Under certain conditions the 
evolution leads to this static solution. The other possible final states correspond to a decompactifying internal sphere,
while the de Sitter space may either expand or contract.

\section{Acknowledgments}

It is a pleasure to thank Indrajit Mitra, Uday Varadarajan, and especially Willy Fischler  for many useful discussions.
This material is based upon work supported by the National Science Foundation under
Grant No. PHY-0071512, and with grant support from the US Navy, Office of Naval Research, Grant Nos. N00014-03-1-0639 and N00014-04-1-0336, Quantum Optics
Initiative.

%



\newpage
\begin{thebibliography}{19}        

\bibitem{Freund:1980xh}
P.~G.~O.~Freund and M.~A.~Rubin,
Phys.\ Lett.\ B {\bf 97}, 233 (1980).

\bibitem{Bousso:2002fi}
R.~Bousso, O.~DeWolfe and R.~C.~Myers,
Found.\ Phys.\  {\bf 33}, 297 (2003)
[arXiv:hep-th/0205080].


\bibitem{DeWolfe:2001nz}
 O.~DeWolfe, D.~Z.~Freedman, S.~S.~Gubser, G.~T.~Horowitz and I.~Mitra,
 Phys.\ Rev.\ D {\bf 65}, 064033 (2002)
 [arXiv:hep-th/0105047].


\bibitem{Contaldi:2004hr}
C.~R.~Contaldi, L.~Kofman and M.~Peloso,
JCAP {\bf 0408}, 007 (2004)
[arXiv:hep-th/0403270].

\bibitem{Kachru:2003aw}
S.~Kachru, R.~Kallosh, A.~Linde and S.~P.~Trivedi,
Phys.\ Rev.\ D {\bf 68}, 046005 (2003)
[arXiv:hep-th/0301240].

\bibitem{Saltman:2004jh}
A.~Saltman and E.~Silverstein,
arXiv:hep-th/0411271.


\bibitem{Giddings:2004vr}
  S.~B.~Giddings and R.~C.~Myers,
  Phys.\ Rev.\ D {\bf 70}, 046005 (2004)
  [arXiv:hep-th/0404220].

\bibitem{Okada:1984cv}
Y.~Okada,
Phys.\ Lett.\ B {\bf 150}, 103 (1985).




\bibitem{Kolb:1986nj}
E.~W.~Kolb,
FERMILAB-PUB-86-138-A



\bibitem{Navarro:2004mm}
  I.~Navarro and J.~Santiago,
  JCAP {\bf 0409}, 005 (2004)
  [arXiv:hep-th/0405173].




\bibitem{Lu:2003dn}
  H.~Lu and J.~F.~Vazquez-Poritz,
  JCAP {\bf 0402}, 004 (2004)
  [arXiv:hep-th/0305250].

\bibitem{Lu:2003iv}
  H.~Lu, C.~N.~Pope and J.~F.~Vazquez-Poritz,
  Nucl.\ Phys.\ B {\bf 709}, 47 (2005)
  [arXiv:hep-th/0307001].

\bibitem{Martin:2003yh}
  J.~Martin, G.~N.~Felder, A.~V.~Frolov, M.~Peloso and L.~Kofman,
  Phys.\ Rev.\ D {\bf 69}, 084017 (2004)
  [arXiv:hep-th/0309001].

\end{thebibliography}
\end{document}